\documentclass[aps,floats,twocolumn,prl,showpacs]{revtex4}               
\usepackage{graphicx,times} 

\newcommand{\wb}{{\mathbf{w}}}    

\def\beq{\begin{equation}}
\def\eeq{\end{equation}}
\def\beqa{\begin{eqnarray}}
\def\eeqa{\end{eqnarray}}
\newcommand{\beqas}{\begin{eqnarray*}}
\newcommand{\eeqas}{\end{eqnarray*}}

\def\sign{\mbox{sign}}

\def\<{\langle}
\def\>{\rangle}

\def\hC{G}

\def\hQ{\hat{Q}}     
\def\hR{\hat{R}} 
\def\eps{\epsilon}          

\begin{document}

\title{Statistical Mechanics
of Learning: A Variational Approach for Real Data} 
\author{D\"orthe Malzahn and Manfred Opper}
\affiliation{Neural Computing Research Group,
School of Engineering and Applied Science,
Aston University,
Birmingham B4 7ET, United Kingdom}
\date{January 10, 2002} 

\begin{abstract}
Using a variational technique,
we generalize the statistical physics approach
of learning from random examples to make it applicable to 
real data. We demonstrate the validity and relevance of our method
by computing approximate estimators
for generalization errors that are based on training data alone.  
\end{abstract}
\pacs{84.35.+i,02.50.-r,05.20.-y,87.18.Sn}           
\maketitle

In recent years, methods of statistical physics
have contributed important insights to the theory
of learning from example data with neural networks and other learning machines
(for recent reviews see e.g. \cite{Eng,Nishi}).
Such systems are often described by models with a large 
number of degrees of freedom which interact by a random energy function where 
the randomness in a learning problem is induced by the data.
Exact solutions for the learning performance
of a great variety of models have been obtained
using tools which were developed for 
the physics of disordered materials, such as the replica trick 
\cite{MePaVi87}. 
All these case studies 
assume simple distributions of the data and 
give invaluable insights into the generic learning behavior. 
However, they do not (and did not intend to) describe real world 
learning experiments where one has the additional problem that the 
distribution of the data is often unknown and idealized assumptions 
about it seem to be too restrictive.
It would be highly desirable if the statistical physics
theory could provide practitioners with tools to
predict and optimize the performance of a learning algorithm.

In this Letter, we will present a step forward in this direction.
We combine the replica approach with a variational approximation
which enables us to deal with more realistic distributions
of the randomness. For models of supervised learning \cite{Eng},
we demonstrate that the theory 
can predict relations between experimentally measurable 
quantities even though details of the data distribution
are unknown. As an application, we will derive approximate expressions
for data averaged performance measures for state of the art
learning algorithms and test them on real data sets. 
We hope that our approach will inspire similar research in related
complex adaptive systems such as, for example,
communication systems \cite{Nishi}.

We demonstrate the basic idea of our approach on a typical
learning scenario where we try to learn an unknown rule which maps
inputs $x\in R^d$ to outputs $y$ from a set $D$ of $m$ example data
pairs $(x_i,y_i)$, $i=1,\ldots,m$. The possible rules are modeled
by a class of real valued functions $f(x)$.
To get a reasonable predictor $\hat{f}(x|D)$ on 
arbitrary novel inputs $x$, a popular learning strategy 
is to balance the goodness of fit on the training data
measured by a training energy $E[f;D] = \sum_{i=1}^m h(f(x_i),y_i)$
with the prior knowledge about the complexity of rules.
In a probabilistic, Bayesian approach to learning \cite{Eng,Nishi}, both ingredients
are combined in a probability distribution 
\beq\label{Gibbs}
\mu_m[f] = \mu[f]\; e^{-E[f;D]}/ Z_m \ ,
\eeq
which is of the form of a Gibbs equilibrium distribution in statistical
physics. It assigns different weights to functions $f$ of being 
responsible for the observed training examples $D$.
$\mu[f]$ encodes the prior knowledge about the plausibility
of different functions. Proper choices of $\mu[f]$ will penalize
functions which fit the data well but are too complex to generalize
properly on novel inputs $x$. Parametric models express $f$ by a set of
parameters $w$, which may
, e.g., be the weights of a neural network.
The distribution $\mu[f]$ is then induced by a distribution over $w$.
Non-parametric models are obtained by assigning
an a priori statistical weight $\mu[f]$ directly over the
space of functions. Predictions for the output $y$ on a novel input $x$
can be made by suitably averaging the function value $f(x)$
weighted by the distribution (\ref{Gibbs}), 
where in the course of learning, when more and more data are observed, 
typically $\mu_m[f]$ will become increasingly concentrated around its mean. 
For continuous outputs (regression), we predict
$y = \hat{f}(x|D) = \<f(x)\>$, and
for binary classification problems, we predict $y = \sign[\<f(x)\>]$,
where angle brackets denote averages over (\ref{Gibbs}).

We will now specialize on models
which are defined by Gaussian prior distributions $\mu[f]$ over functions.
Assuming a zero mean, they are fully specified 
by the correlation {\em kernel} $K(x,x') \doteq 
\int d\mu[f] \; \{f(x) f(x')\}$ which must be supplied by the user.
It encodes a priori assumptions about the typical variability 
of model functions $f$ with the input $x$.
Such {\em Gaussian process} models (GP) have attracted considerable 
attention in recent years as they represent a flexible and widely 
applicable concept \cite{whaba,Nealbook,WiRa96,Bialek}. 
GP models can be understood as a limit of Bayesian 
feed-forward neural networks when the
number of hidden units grows to infinity \cite{Nealbook}. 
Non-probabilistic ``kernel machines'' 
such as the celebrated {\em support vector machines} 
(SVMs) \cite{Kernel} result from GP models by taking appropriate limits.
GP also form a basis for field theoretic approaches to density
estimation \cite{Bialek}.
For all GP models, the mean predictions are expressed  as
a linear combination of kernel functions centered at 
the input data    
$\<f(x)\> = \sum_{j=1}^m \alpha_j K(x,x_j)$, where
the $\alpha_j$'s are certain Gibbs averages which are 
independent on the point $x$ \cite{OpWi00,Nealbook,WiRa96}.

We study the typical learning performance of this
kernel approach by averaging over
different drawings of training data sets $D$
where all examples $(x_i,y_i)$ are generated 
independently from the same distribution $p(y,x) = p(y|x) p(x)$. We use  
the replica approach for computing 
the averaged free energy $F= [-\ln Z_m]_D$ which serves as a generating
function for useful data averages. 
Here, we denote data averages by square brackets. We get
$F =-  
\lim_{n\to 0}\frac{\partial \ln [Z_m^n]_D}{\partial n}$.
To facilitate subsequent calculations, we use 
a {\em grand canonical} formulation
where the number of examples $m$ is only fixed on average by
a chemical potential $\mu$. An elementary calculation
which uses the independence of the data,
yields the grand canonical partition function 
for the n times replicated system
\beq\label{grandcan}
\Xi_n(\mu) \doteq \sum_{m=0}^{\infty} \frac{e^{\mu m}}{m!}[Z_m^n]_D =
\int \prod_{a=1}^n d\mu[f_a]\; \exp [- H]
\eeq
in terms of a Hamiltonian $H = \left[{\cal{H}}(\{f_a\},x)\right]_x$
which is the average of a {\em purely local} 
Hamiltonian density
\beq\label{hamil}
{\cal{H}}(\{f_a\},x) =
- e^{\mu} \left[\exp\left\{-\sum_{a=1}^n h(f_a (x), y)\right\}\right]_{(y|x)}
\ .
\eeq
Here, the square brackets $[\ldots]_x$ and $[\ldots]_{y|x}$
denote expectations with respect to the distribution of inputs $x$ and
with respect to the conditional distribution of outputs $y$ 
given the inputs.
The chemical potential $\mu$ is adjusted such that 
$
m = \frac{\partial \ln \Xi_n(\mu)}{\partial \mu}
$
for all $n$ which gives 
the simple result $\mu = \ln m$ in the limit $n\to 0$.
For sufficiently large $m$, we replace
the sum over $m$ in Eq. (\ref{grandcan}) by its dominating term and
recover the original (canonical) free energy
as
$
F = - \lim_{n\to 0}\frac{\partial \ln \Xi_n(\ln m)}{\partial n}.
$

Rather than evaluating (\ref{grandcan}) for simple 
and artificial distributions $p(x,y)$
in the limit of high input dimensionality,
we resort to a variational approximation which can adapt to
practically relevant situations. 
A similar approach was found to be useful in studying low dimensional 
disordered systems such as polymers or interfaces in 
random media \cite{rmedia}. We approximate $H$  
by a trial replica Hamiltonian $H_0$ which minimizes the 
variational bound (see e.g. \cite{feynman}) 
\beq\label{varbound}
-\ln \Xi_n(\mu) \leq -\ln \int \prod_{a=1}^n d\mu[f_a] 
\; e^{- H_0} + \< H - H_0\>_0 \ .
\eeq
The brackets $\<\ldots\>_0$ denote an average with respect to
the distribution induced by $\prod_{a=1}^n\mu[f_a]e^{-H_0}$. 
Equation (\ref{varbound}) is the leading
term in a systematic perturbation expansion of the free energy with respect
to the difference $H - H_0$.
For Gaussian measures $\mu[f]$, a local trial Hamiltonian 
of the form  
\beq\label{trial}
H_0 = \left[ \sum_{a\le b}\hQ_{a b}(x) f_a(x) f_b(x)
+ \sum_a \hR_a(x) f_a(x)\right]_x
\eeq
is an appropriate choice. The most important feature of 
(\ref{trial}) is the explicit
dependence of the variational parameters $\hQ_{ab}(x)$ and $\hR_a(x)$
on the input variable $x$ . This enables us to take the nonuniformity of
realistic input densities into account. 
The resulting Gaussian approximation is expected to become
asymptotically exact for training energies  
$h(f,y)$ that are smooth functions of $f$, when the Gibbs
distribution (\ref{Gibbs}) becomes increasingly concentrated around its mean 
for large $m$.  

Expressing the averaged Hamiltonian $\<H\>_0$ by 
local order parameter fields 
$R_a(x)=\<f_a(x)\>_0$ and $Q_{ab}(x,x)$
[the latter being a special case of the two point
function $Q_{ab}(x,x') = \<f_a(x) f_b(x')\>_0$], a 
straightforward variation yields
\beq\label{ordpareq}
\frac{d \<{\cal{H}}\>_0}{ d R_a(x)} = \hR_a (x)\ ;\qquad 
\frac{d \<{\cal{H}}\>_0}{d Q_{ab}(x,x)} = \hQ_{ab}(x)
\ .
\eeq  
Assuming replica symmetry, i.e., $R_a(x) = R(x)$, 
$Q_{ab}(x,x') = Q(x,x')$
for $a\neq b$ and $Q_{aa}(x,x') = Q_0(x,x')$,
the order parameter fields have a simple physical meaning in terms 
of averages over example data sets $D$. We have
$Q(x,x') = [\<f(x)\>\<f(x')\>]_D$ and 
$Q_0(x,x') = [\<f(x) f(x')\>]_D$. 
$\hC(x,x') = [\<f(x) f(x')\> - \<f(x)\>\<f(x')\>]_D$ is the average posterior
correlation function. $R(x) = [\<f(x)\>]_D$ is the
bias of the predictor, whereas $V(x,x')=Q(x,x') - R(x) R(x')$ 
is its data covariance.
A direct calculation of order parameters from (\ref{trial})
expresses these in terms of the variational parameters as
$R(x)= - [\hC(x,x') \hR(x')]_{x'}$ 
and $V(x,x') =
- [\hC(x,x'') \hC(x',x'') \hQ(x'')]_{x''}$. Finally, $\hC$ 
is found as the operator inverse 
$\hC = \left( K^{-1} + u \right)^{-1}$,
where $K$ is the kernel integral operator and 
$u(x,x') = p(x)(\hQ_0(x) - \hQ(x))\delta(x - x')$.

There are various ways to make use of our variational framework.
We can use the probability measure defined by the 
trial Hamiltonian (\ref{trial}) in order to compute
the average case performance  
of the learning algorithm at test points $(x,y)$ not contained 
in the data set $D$.
For example, the data average of the  
(mean square) prediction error 
$\varepsilon_2(D) = [(\<f(x)\> - y)^2]_{x,y}$ is 
$[\varepsilon_2(D)]_D = [(R(x) - y)^2 + V(x,x)]_{x,y}$. Its sample 
fluctuations are
$
[\varepsilon_2(D)^2]_{D}-[\varepsilon(D)]^2_{D} =
[4(R(x)-y)(R(x')-y')V(x,x') + 2 V^2(x,x')]_{x,x',y,y'}
$. 
Similar to previous studies in the statistical mechanics
of learning \cite{Eng,Nishi}, we could 
compute explicit learning curves 
from the variational equations, when the distribution of data
is given. In general, explicit
analytical solutions are possible
for simple distributions,
or in the asymptotic limit when the number of data grows large.
We will give examples of such results elsewhere. 
Since in practice, however, data distributions are usually unknown, 
a more important application of our approach is in the 
derivation of explicit {\em relations} between
data averaged observables, which will hold (within the framework
of our approximation) for {\em any} such distribution. 
Such relations can help to estimate the performance of an algorithm
on novel data which were not contained in the training set $D$.

To demonstrate this idea, 
we consider the problem of finding
empirical estimates for 
generalization errors on novel test data $(x,y)$. These estimates
should be computable from the training data $D$ only.
Since in many real applications, test errors 
may be measured with an error function
different from the original training energy $h$, we 
consider general error functions    
$L(\<f(x)\>,x,y)$. Obviously, 
using the naive approximation $\frac{1}{m}\sum_{i=1}^m L(\<f_i\>,x_i,y_i)$
to estimate the expected error $\varepsilon_L(D) = [L(\<f(x)\>,x,y)]_{x,y}$, 
where
$f_i \doteq f(x_i)$, will give an optimistically biased estimate 
in most cases. In order to compute better, {\em unbiased} estimates
for test errors which are based on the training data, we use the 
general result\beqa\label{empire}
\frac{1}{m}\left[\sum_{i=1}^m g(\<F_1(f_i)\>,\ldots,
\<F_k(f_i)\>,x_i,y_i)\right]_D = \\
\nonumber
\int Dz\; [g(\<F_1(\phi)\>_{\phi},\ldots,
\<F_k(\phi)\>_{\phi},x,y)]_{x,y}
\eeqa
where $Dz=e^{-z^2/2}dz/\sqrt{2\pi}$ and  $\<\ldots\>_{\phi}$ denotes an
 expectation with respect to the distribution
\beq\label{phidist}
P(\phi) = \frac{\exp\left[- h(\phi,y) - \frac{
(R(x) + z \sqrt{V(x,x)} - \phi)^2}{2\hC(x,x)}\right]}
{\sqrt{2\pi \hC(x,x)}\; Z(x,y,z)} 
\eeq
with norm $Z(x,y,z)$. Eq. (\ref{empire}) is easily
proved within our variational framework and holds for arbitrary functions
$g$, $F_1,\ldots,F_k$ \cite{Fluct}. 
To compute the desired unbiased estimates, we try to find a set
of functions $g$ and $F_1,\ldots F_k$ such that the right 
hand side of (\ref{empire})
can be rewritten as the data average of $\varepsilon_L(D)$. 

\begin{figure}[t]
  \centerline{
      \scalebox{0.28}[0.27]{\includegraphics{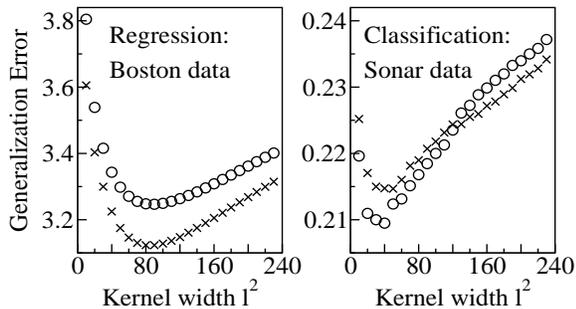}}}
  \caption{\label{fig1}
Model selection based on Eq. (\ref{lossest}) for a regression problem (Boston 
housing)
and a binary classification problem (Sonar) using the kernel 
$K(x,x')=\exp(-||x-x'||^2/l^2)$. The right hand side of Eq. (\ref{lossest}) 
is an
empirical estimate (crosses) which is calculated
only on {\em training} data. It gives a good account of the
generalization error (circles) which was calculated on {\em  test}
data. 
}
\end{figure}

We will demonstrate this idea for the GP model with mean square
training error 
$h_2(f(x),y) = \frac{1}{2\sigma^2}(f(x) - y)^2$.
This model finds widespread applications
\cite{whaba,WiRa96} and has the advantage that the computations of Gibbs
averages (\ref{Gibbs}) such as means and correlation functions  
can be performed analytically in closed form.
Nevertheless, the analysis of the data average is nontrivial, because
averaging leads to a non Gaussian model in replica space
which makes the application of our variational approximation still necessary. 
Using $F_1(f) = f$ and $F_2(f) = f^2$ in Eq. (\ref{empire}) 
as well as 
$[L(\<f(x)\>,x,y)]_{D} = \int Dz \; [L(R(x) + z \sqrt{V(x,x)},x,y)]_{x,y}$ 
yields
\beq\label{lossest}
[\varepsilon_L(D)]_D =
\frac{1}{m}\sum_{i=1}^m \left[L\left(\frac{\sigma^2 \<f_i\> - \sigma^2_i y_i}
{\sigma^2  - 
\sigma^2_i},x_i,y_i\right)\right]_{D}  \ ,  
\eeq
where
$\sigma^2_i = \<f^2_i\> - \<f_i\>^2$.
Generalizations to other GP models and support vector machines 
are possible and will be given elsewhere. We present tests of 
Eq. (\ref{lossest})  
on regression and binary classification problems. In both cases
$\<f(x)\>$ is computed from the squared error $h_2$ with 
$\sigma^2=0.01$ (Boston data) and $\sigma^2=0.1$ (Sonar data), respectively.   
The left panel of Fig. \ref{fig1} compares the generalization errors
and their estimates for the publicly available {\em Boston housing}
regression data set \cite{datasets}
using the error function $L(\<f(x)\>,x,y)= |\<f(x)\> - y|^k$ for 
$k=1$, $m = 50$ 
and different widths $l^2$ of the kernel $K(x,x')=\exp(-||x-x'||^2/l^2)$. 
The results (which are averaged over 20 splits of the entire data set into 
training and test examples) suggest that our estimators might be well used 
for {\em model selection}, i.e., for finding the optimal kernel parameters
with smallest generalization error. 
It is interesting to note that the case $k=2$ leads to the exact 
{\em leave one out estimator} for the square error described in \cite{whaba}. 
The right panel of Fig. \ref{fig1} shows corresponding results
for binary classification ($y =\pm 1$) on the {\em Sonar data}, where 
now the generalization error is the average 
fraction of misclassified test points, i.e.,
$L(\<f(x)\>,x,y)= \Theta(-y\<f(x)\>)$, and
$\Theta(x)$ is the unit step function.

\begin{figure}[t]
\centerline{
      \scalebox{0.28}[0.27]{\includegraphics{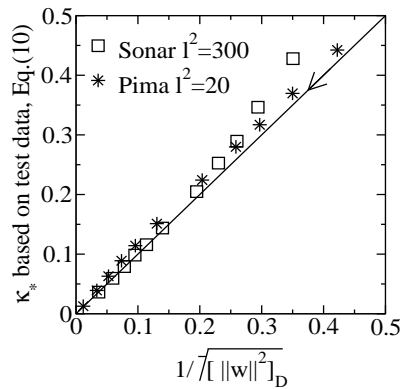}}}
\caption{\label{fig2}
Verification of Eq. (\ref{margin}) on two benchmark data sets for binary classification
with input dimension $d=7$ (Pima Indian diabetes, stars) and 
$d=60$ (Sonar data, squares). 
The arrow points into the direction
of increasing number $m$ of training data with
$m=4-7$ in steps of 1, $m=9-24$ in steps of 5 and $m=33,50,100$.
}
\end{figure}

In the following, we will test the validity of 
our variational replica approach for a case where
the combination of a non-smooth training energy with a subtle 
singular limit might not suggest immediately  
that the trial Gaussian distribution (\ref{trial}) is a good
approximation.
We consider SVM classifiers \cite{Kernel} which
can be understood as generalizations of   
single layer neural networks which allow for {\em nonlinear}
separation between classes. Expanding the positive definite SVM kernel
$K(x,x') = \sum_k \psi_k(x)\psi_k(x')$
in a set of implicit features
$\psi_k(x)$, the SVM output can be written as $y =\sign[f(x)]$ where
$f(x) = \sum_{k}  w_k \psi_k(x)$. 
The vector $\wb$ of 
SVM weights $w_i$ is determined by minimizing its
length $||\wb||$ under the condition that the ``hard margin''
training energy $h_{HM} =0 $. The latter is defined as 
$h_{HM}(f(x),y) = 0$ if $y f(x) \geq 1$ and $h_{HM} = \infty$ otherwise. 
This SVM model is contained within the present GP
framework \cite{OpWi00} by introducing a Gaussian prior 
distribution of variance $\eps$ independently for each weight $w_k$. 
In the the limit
$\eps\to 0$ the posterior Gibbs distribution becomes concentrated at
the minimal length weight vector of the SVM.
The Gaussian distribution over weights
is equivalent to a Gaussian process over functions $f$
with correlation kernel $K_{\eps}(x,x') = \eps K(x,x')$. 
We test our method on the SVM margin 
$\kappa(D) = 1/||\wb||$ 
between positive and negative labeled examples
which plays an important role as an indicator for good
generalization ability of SVMs \cite{Kernel}. It can be shown that  
$\kappa(D)$ is related to the free energy as 
$1/\kappa^2_* \doteq [1/\kappa^2(D)]_D = 2 \lim_{\eps\to 0} \eps F_{\eps}$. Using our variational 
approximation we get
\beq\label{margin}
1/\kappa^2_* 
=\left[\frac{m \sqrt{V(x,x)}}{\chi(x,x)} 
\int\limits_{-\infty}^{\Delta(x,y)} \!\! Dz
\left(\Delta(x,y) - z \right)\right]_{x,y}
\eeq
where $\Delta(x,y)=\frac{1-y R(x)}{\sqrt{V(x,x)}}$ and
$\chi(x,x') = \lim_{\eps\to 0} \eps^{-1} \hC_{\eps}(x,x')$
is a response function which can be computed using the SVM algorithm. 
Equation (\ref{margin}) relates
the averaged inverse squared margin on the training data to functions of
the SVM's {\em bias}, its {\em variance} and the response function 
$\chi$ on novel test data $(x,y)$. It generalizes 
Gardner's famous result \cite{Gardner} for
the optimal stability of linear perceptrons
to SVMs under general data distributions.
Figure \ref{fig2} compares $\kappa_*$ as computed from its basic definition,
with our prediction given by Eq. (\ref{margin}) for two 
publicly available classification data sets \cite{datasets}. 
We used the kernel $K(x,x')=\exp(-||x-x'||^2/l^2)$ \cite{RBF}. 
Figure \ref{fig2} is obtained for suboptimal choices of 
the correlation length $l$ and 
different sizes $m$ of the training data sets $D$. For each $m$ 
we have performed an average over 20 splits of the entire data set  
into training and test examples. 
Our variational Gaussian approximation improves with growing training size
$m$ and higher data dimensionality.
The accuracy of our result (\ref{margin}) 
improves further, if the kernel width $l^2$ is optimized to achieve
small generalization errors.

The two examples presented so far are only a small selection of
the possible applications of our approach. Obvious future extensions
will include an assessment of the reliability of estimators 
Eq. (\ref{lossest}) for model selection
by taking their sample fluctuations into account. Also
alternative, Bayesian criteria which 
use the minimization of the free energy for model selection   
fit naturally into the statistical physics framework. Generalizing 
the variational approach to models with statistically
{\em dependent} data such as e.g. (hidden) Markov processes, which are
relevant for time series prediction, 
and to non Gaussian priors and trial Hamiltonians, will extend
its applicability to the wider field of modern probabilistic data modeling.
Finally, it will be necessary to estimate and improve the accuracy
of our approximations. This can be done by computing perturbative 
corrections to the variational free energy (\ref{varbound}), but also
in a nonperturbative framework by including the possibility of
{\em replica symmetry breaking}. While the latter phenomenon
is not expected to be relevant for most of the interesting kernel models
with convex error functions, applications of the method in other
fields like combinatorial optimization might definitely benefit
from this extension.

{\bf Acknowledgement:} This work was supported by EPSRC grant GR/M81601.


\begin{thebibliography}{99}
\bibitem{Fluct} Similarly, one can compute sample fluctuations.
\bibitem{RBF} One can model decision surfaces by 
polynomials of arbitrary order.
\end{thebibliography}
\end{document}